\documentclass[pra,twocolumn,showpacs]{revtex4}
\usepackage{hyperref}
\usepackage{graphicx,psfrag}
\usepackage{amsmath,amssymb,bm}

\newcommand{\ket}[1]{|#1\rangle}
\newcommand{\bra}[1]{\langle#1|}
\newcommand{\braket}[2]{\langle#1|#2\rangle}
\newcommand{\im}{{\rm i}}

\newcommand{\abs}[1]{|#1|}
\newcommand{\iden}{\bm{1}}

\begin{document}

\title{Robust generation of entanglement between two cavities mediated by short interactions with an atom} \date{\today}

\author{Daniel E. Browne}
\author{Martin B. Plenio}
\affiliation{Blackett Laboratory, Imperial College London, London SW7 2BW, UK.}
\pacs{03.67.-a, 42.50.-p, 03.65.Ud}
\begin{abstract}
Current methods of preparing maximally entangled states in the
laboratory need an extremely accurate control of interaction
times, requiring sophisticated experimental techniques.
   Here, we show that such precise control is not
necessary when one utilizes short or weak interactions followed by
measurements. We present a scheme for the probabilistic generation
of Bell states in a pair of cavities, after each has interacted
briefly with an atom. The advantage of the scheme , as
compared to present schemes,  is its low sensitivity
to the exact values of experimental parameters such as atomic
velocity and coupling strength, in fact, for a large range of
parameters the fidelity of the Bell states generated remains close
to unity.
\end{abstract}

\maketitle

\section{Introduction}

In recent years  entanglement, a fundamental feature of many-body
quantum mechanical systems, has come to be seen as a useful
resource for many tasks in quantum information processing
\cite{nielsenchuang,wernerinalberbook,pleniovedralreview}.
In particular, certain states have been identified, which can be
directly used for many such tasks. Foremost among these are the
so-called Bell states, also known in the literature as EPR states,
which exhibit the maximum entanglement which a two qubit system
can possess. Such states are the fundamental ingredient of in a
variety of quantum information processing protocols (see
\cite{nielsenchuang,wernerinalberbook,pleniovedralreview} for
details).
Bell states are also of relevance to the study of the foundations
of quantum mechanics. It was shown by Bell \cite{bell} that these
states possess correlations which are incompatible with a local
realistic theory (see \cite{someone} for a review). Thus, a
loophole-free demonstration of such correlations would have
profound implications for our understanding of the physical world.
For these reasons, then, it is important to be able to generate
Bell states in the laboratory, and many experimental proposals
have been put forward and realized in both optical
\cite{wu1950,kwiatdownconv,lamasstim} and atomic systems
\cite{rowebell,frybell}.

This paper contains a proposal for a scheme utilizing the
techniques of cavity QED to generate Bell states between the modes
of two spatially separated cavities. Rauschenbeutel and coworkers
have generated Bell states between two modes in a single cavity
using a Rydberg atom \cite{rauschen01} coherently interacting with
each mode in turn. This scheme could be adapted in a
straightforward way to generate such states in spatially separated
cavity modes.
Cabrillo {\em et al}\/\ have proposed a non-deterministic scheme
to generate Bell entangled states between atoms conditionally, by
driving them with a weak laser pulse, and subsequently detecting a
photon spontaneously emitted by one of the atoms
\cite{cabrillo99}. Protsenko and coworkers have then shown that
this scheme can be adapted to implement quantum logic gates
\cite{protsenko}. At the same time Plenio {\em et al} have
developed schemes to entangle atoms inside optical cavities and
between atoms in different cavities via spontaneous decay
\cite{pleniohuelga99,bose99tele,janeplenio}. Various other
conditional schemes and deterministic schemes have been proposed
which employ the detection of photons to generate entanglement and
perform quantum computation \cite{knilllaflamme,kok02,fiurasek02}.
In schemes such as these, however, a very precise control of the
experimental parameters is required. Such a precise control of
experimental parameters is often difficult to achieve. Generally
it is desirable to devise schemes in which the requirements on the
experimental control are as small as possible. Such schemes would
then be far more robust to errors and would generally lead to much
improved fidelities of the generated states.

In this paper, we present a novel proposal for the robust
experimental generation of a single photon entangled Bell state
$\ket{\Psi^+}=\sqrt{\frac{1}{2}}\left(\ket{0}_A\ket{1}_B+\ket{1}_A\ket{0}_B\right)$
in the modes of two spatially separated cavities, labeled $A$ and
$B$. The cavity modes do not interact directly, but the generation
of the entangled state is mediated by a resonant atom. This atom,
prepared in its excited state, passes through the two cavities,
prepared in their vacuum state, and interacts with each for the
same short effective interaction time. The internal state of the
atom is then measured. Depending on the outcome of this
measurement, either a high fidelity Bell state has been generated,
or the cavities have returned to the vacuum state, and are ready
for the process to be repeated with a fresh atom.
The scheme has the advantage that provided the interaction times
between the atom and both cavities are the same, which can be
attained if a well-collimated atomic beam is utilized, the actual
value of this interaction time can vary within a large range of
values, without greatly affecting the fidelity of the Bell state
produced. This would reduce the need for velocity selection of the
atoms and  for the sophisticated timing and feedback required to
tune the interaction times in other experiments.
Additionally, the interaction times, which this scheme would
utilize, can be much shorter than in non-conditional schemes, which
would reduce the effects of decoherence during the entanglement
generation process.

In section \ref{scheme}, we introduce the proposed scheme in
detail.  We begin by  discussing the
scheme in the limit where the effective duration of the
atom-cavity interactions is very small, and the same in both
cavities. We then discuss the fidelity of the state generated by a
successful run of the scheme with finite interaction times and
calculate the probability that a single run of the scheme is
successful. We find that, in a wide range of interaction times,
the probability of success is suitably high, and the fidelity of
the generated state is close to unity.
In section \ref{practical} we discuss practical aspects of
implementing our scheme in three subsections. In the subsection
\ref{implementation}, we discuss specific physical systems with
which the scheme could be implemented.
In the  two subsections which follow this  we consider two practical
aspects of the scheme which would pose particular requirements to
any implementation of the scheme. In  subsection \ref{collim} we
look at the effect on the scheme of inaccurate control of the path
of the atom through the cavities due to poor collimation of the
atomic beam. This would lead to non-equal interaction times in the
two  cavities, potentially reducing the fidelity of the state
generated. In  subsection \ref{detection}, we discuss how
employing detectors with less then perfect efficiency would affect
the scheme.

\section{The proposed scheme}\label{scheme}

\begin{figure}
\includegraphics[scale=0.3]{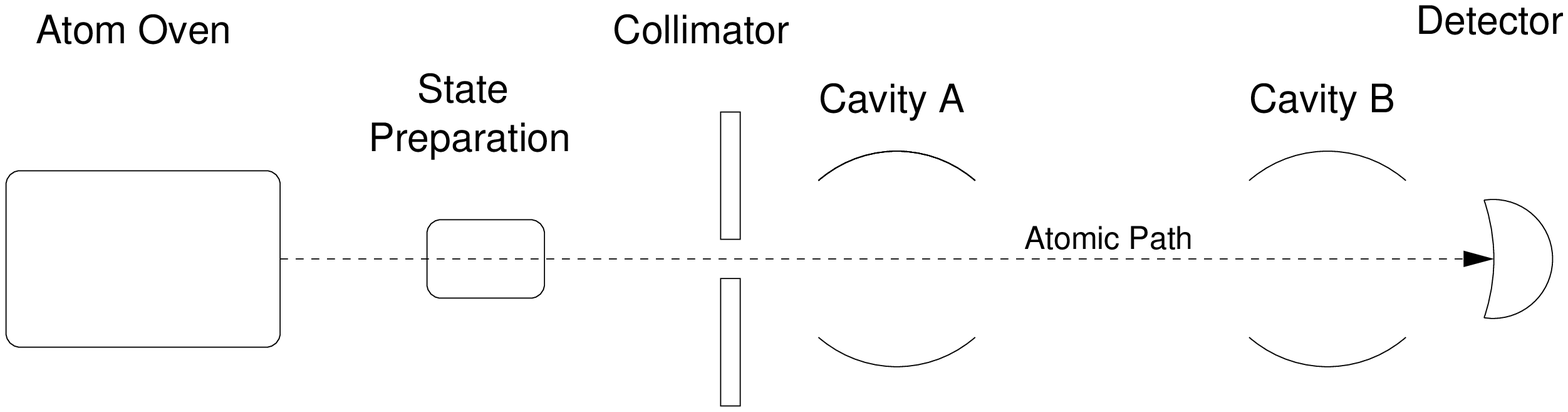}
\caption{\label{fig:layout} A schematic diagram of the layout of the scheme.}
\end{figure}

In this section, we will describe the proposal which is the main
focus of this paper, a scheme to generate the entangled Bell state
$\ket{\Psi^+}$ in two spatially separated microwave cavities.
Two identical cavities, labeled A and B,  are positioned as in
figure \ref{fig:layout}, such that their axes are parallel, and
their centers in alignment. The cavities must be prepared in their
vacuum state. This can be achieved, for example,  in  microwave
cavities by cooling to low temperatures.
Two-level atoms, whose transition between the ground state $\ket{g}$ and excited state  $\ket{e}$ is resonant with modes of the cavities, are used to mediate the generation of the entangled state in the cavities.
The atoms are prepared in their excited state, $\ket{e}$. They are then passed, one at a time through the center of both cavities, such that the effective interaction time with each cavity is much less than the period of a one-photon Rabi oscillation. Upon leaving the second cavity and before the next atom enters the cavities, the internal state of the atom leaving the cavities is measured. For brevity in this paper, let us refer to the process of a single atom passing through both cavities and being measured as a ``run'' of the scheme.
The outcome of the measurement is non-deterministic. If  the excited state $\ket{e}$ is detected, then no entangled state has been generated in the cavities, we thus refer to a run which ends in such a measurement as a ``null run''
After a null run, the cavity is reset to its initial state, so a further run can be implemented immediately.
If the ground state $\ket{g}$ of the atom is detected at the end of a run, a state is generated in the cavities, which, so long as certain conditions are fulfilled, as described below, is very close to the Bell state $\ket{\Psi^+}$. The process is complete at this point, and we shall thus label this a ``successful run''.

Before we describe the scheme in detail, let us summarize how the resonant interaction between atom and cavity is described.
Consider a two-level atom with energy levels $\ket{g}$ and $\ket{e}$, separated by a transition energy resonant with a mode of the cavity. The atom is
situated at position $\bm{r}$ within the cavity, whose modes are described by field operators $a$ and $a^\dag$.  Since the atomic transition is resonant with the cavity mode, the internal atomic state interacts with this mode via the following Hamiltonian, in the interaction picture, where the rotating wave approximation has been employed,
\begin{equation}\label{atomcavham}
H=\hbar g_0 u(\bm{r})\bigl( a^{\dag} \ket{g}\bra{e}+a\ket{e}\bra{g}\bigr).
\end{equation}
Here $g_0$ describes the strength of the coupling. When the atom is situated in the center of the cavity, the one-photon Rabi frequency is $g_0/\pi$. The spatial variation of the cavity field is contained in the mode function $u(\bm{r})$. If we
consider an atom moving through the cavity on a classical path $\bm{r}(t)$, the mode function can then be written as a function of time $u(t)$.  Since all the time-dependence in $H$ is contained in $u(t)$, the unitary evolution of the system can be written as follows
\begin{equation}
U=\exp\left[- \frac{\im \left( \int_{t_0}^t u(t')dt'\right)
\hat H}{\hbar}\right],
\end{equation}
where $\hat H= H/u(\bm{r})$ is the interaction Hamiltonian for an atom situated in the center of the cavity.  In this article, we will consider cases where the atom travels on a straight path through the cavity, and can thus introduce the total effective interaction time  $\tau=\int_{-\infty}^\infty u(t') dt'$ to characterize the interaction.
%, where to simplify its calculation, we have extended the integration limts to plus and minus infinity, which is a reasonable approximation for an atom travelling on a stright path, since  the coupling constant $u(\bm{r})$ is negligible when the atom is far from the centre of the cavity. 
The complete unitary evolution of of the system due to the interaction $U_{\text{total}}$ is then
\begin{equation}
U_{\text{total}}=\exp\left[- \frac{\im \tau 
\hat H}{\hbar}\right].
\end{equation}

In our scheme, the atom passes through and interacts with two cavities in turn, labeled $A$ and $B$.
Let us label the interaction Hamiltonians for the interaction between the atom and each cavity $H_A$ and $H_B$ respectively. These take the same form as $H$ in equation \eqref{atomcavham} for each respective cavity. Analogously let us label the interaction Hamiltonians for the atom in the center of each cavity $\hat H_A$ and $\hat H_B$. The cavities are assumed to be identical and thus the coupling constant $g_0$ is the same for both.
Let us consider how the state of the system will evolve if a resonant atom, initially in its excited state, passes through and interacts with the cavities, one after the other. If the cavities are initially in some pure state $\ket{\psi_{\text{cav}}}$. The initial state of the atom-cavities system is thus $\ket{e}\ket{\psi_{\text{cav}}}$.
The atom is passed through both cavities such that the effective interaction times are $\tau_A$ and $\tau_B$.
After these interactions, the quantum state of the system has
undergone a unitary evolution described by the operator $U_{AB}$
\begin{equation}
    U_{AB}=e^{-\im g_0 \tau_B\hat H_B}e^{-\im g_0 \tau_A\hat H_A} .
\end{equation}

In the scheme we propose, the cavities are aligned as shown in
figure \ref{fig:layout} and the atom passes along the straight
line through the center of both cavities at constant velocity.
This means that the effective interaction times between the atom
and each cavity will be equal. It may, of course, be difficult to
control the path of the atom with sufficient accuracy that the
interaction times are exactly equal, and the effect of this is
discussed in section \ref{collim}. For now, however, let us assume
that $\tau_A$ and $\tau_B$ are equal, and write them both $\tau$.
In the limit when the effective interaction times are very small, i.e. when $g_0\tau\ll1$, $U_{AB}$ can be expanded to the first order in $g_0\tau$, and takes the following form
\begin{equation}
\begin{split}
U_{AB}&\approx\ \iden - \im \frac{H_{A}\tau}{\hbar}- \im \frac{H_{B}\tau}{\hbar}\\
&=\iden - \im g_0\tau\biggl[(a_{A}+a_B) \ket{e}\bra{g}+(a^\dag_{A}+a^\dag_B) \ket{e}\bra{g}\biggr].
\end{split}
\end{equation}

The state of the system after the atom has left the second cavity, is $\ket{\psi}=U_{AB}\ket{\psi_{\text{init}}}$, in this limit
\begin{equation}\label{gencavoutcome}
\ket{\psi}\approx \ket{e}\ket{\psi_{\text{cav}}}-\im g_0\tau\ket{g}(a^\dag_{A}+a^\dag_B)\ket{\psi_{\text{cav}}}.
\end{equation}

When the state of the atom is now measured, the cavity modes are projected into one of two states, depending on the measurement outcome.
If $\ket{e}$ is detected, the cavity returns to its initial state.
This is important for a non-deterministic process, because it
means that it can be repeated immediately from the same starting
conditions. If the ground state $\ket{g}$ is detected, the cavity
modes are now in the state, neglecting normalization,
$(a^\dag_{A}+a^\dag_B)\ket{\psi_{\text{cav}}}$.
Thus, if the cavities are initially in the vacuum state, the state generated in the cavities would be $(a^\dag_{A}+a^\dag_B)\ket{0}_A\ket{0}_B=\ket{1}_A\ket{0}_B+\ket{0}_A\ket{1}_B$, which, when normalized is the Bell state $\ket{\Psi^+}$ introduced above.

If, following a successful run, one were to repeat the scheme
immediately and carry on until $n$ atoms had been detected in the
ground state, in the limit that  $g_0\tau$ is small, the state
generated would have the form
$(a^\dag_{A}+a^\dag_B)^n\ket{\psi_{\text{cav}}}$, or, if the
cavities are initially in their vacuum state,
$(a^\dag_{A}+a^\dag_B)^n\ket{0}_A\ket{0}_B$. This is equivalent
to the state produced when a $n$-photon Fock state and a vacuum
state are incident together on a 50:50 beam splitter. However,
numerical results have suggested that, the fidelity of states
produced via this method would decrease swiftly with increasing
$n$. This is due to two reasons, firstly, the short interaction
time approximation becomes worse when higher photon numbers are
present in the cavities since the timescale of the interactions
is faster (the Rabi frequency scales with $\sqrt{n+1}$), secondly,
when more than one photon is in the cavities, a null run, the
measurement of the atom leaving the cavities to be in its excited
state, does not reset the state of the cavities to the state
before the run, so, as more repetitions are made, the fidelity of
the final state gets worse and worse. For these reasons, this
 does not appear to be  a good scheme for
the generation of such states.

However, as we will show below, the generation of single-photon
Bell states is not affected by these problems. Firstly, we find
that the fidelity of the generated state remains close to unity
for values of $g_0\tau$ much greater than the above approximation
would be valid. Secondly, in this case, null runs reset the state
of the cavities to the vacuum state exactly, so the fidelity of
the state generated is unaffected by the number of runs which were
required.
In the limit that $g_0\tau$ tends to zero, the probability of the
detector measuring a ground state is approximately $2(g_0\tau)^2$,
which, in this limit, is vanishingly small. In a practical scheme,
one would need to work in a parameter range, where the probability
of success was high enough that few repetitions would be required
to achieve a successful run. This is the case for higher values of
$g_0\tau$, where the above approximation would no longer be valid.
Fortunately, starting with the simple pure state
$\ket{e}\ket{0}_A\ket{0}_B$, it is straightforward to solve the
Schr\"odinger equation and calculate exactly the state of the
system after the interactions have taken place. This state
$\ket{\psi}$, under the assumption that the interaction times
$\tau$ are exactly equal is,
\begin{equation}
\begin{split}
    \ket{\psi}=&\cos^2(g_0\tau)\ket{e}\ket{0}_A\ket{0}_B -
    \im\cos(g_0\tau)\sin(g_0 \tau)\ket{g}\ket{0}_A\ket{1}_B\\
&\qquad -\im\sin(g_0\tau)\ket{g}\ket{1}_A\ket{0}_B.
\end{split}
\end{equation}

Let us consider a measurement of the atom's state. If
the excited state is detected the
state of the cavities is projected back to the vacuum state,
independent of $\tau$, as in the approximate case. This resets the cavities to their initial
state, so the process can be immediately repeated with a fresh
atom.

\begin{figure}
\includegraphics{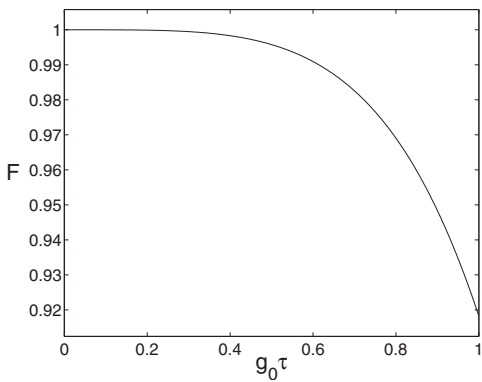}
\caption{\label{fig:idealfid} The fidelity of the Bell state
generated in the cavity modes after a successful run, plotted
against $g_0\tau$ with values from 0 to 1. Note that the values of
$g_0\tau$ corresponding to a full Rabi oscillation in each cavity
is $g_0\tau=\pi$.}
\end{figure}

If the ground state is detected, the following entangled state is
generated in the cavity,
\begin{equation}
\ket{\psi_{\text{cav}}}=\cos(g_0 \tau)\ket{0}_A\ket{1}_B+\ket{1}_A\ket{0}_B,
\end{equation}
 where normalization has been omitted. As $g_0\tau$
approaches zero, this tends to the desired state $\ket{\Psi^+}$.
One can quantify how close this state generated is to
$\ket{\Psi^+}$ in terms of the fidelity
$F=\abs{\braket{\Psi^+}{\psi_{\text{cav}}}}^2$.
\begin{figure}
\includegraphics{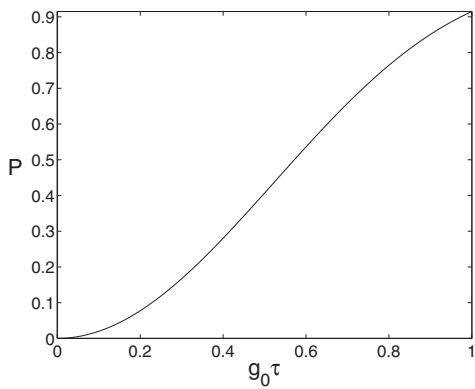}
\caption{\label{fig:idealprob} The probability that a single run will lead to the detection of the atom in its ground state, and thus the successful generation of a Bell state, plotted for values of $g_0\tau$ from 0 to 1.}
\end{figure}
\begin{equation}\label{eq:idealfid}
F=\frac{1}{2}+\frac{\cos(g_0\tau)}
{\cos^2(g_0\tau)+1}=1-\frac{(g_0\tau)^4}{16}+
O\bigl((g_0\tau)^6\bigr)
\end{equation}
This is plotted in figure \ref{fig:idealfid}. We see that the fidelity remains very close to unity for a surprisingly large range of $g_0\tau$. For example, the fidelity remains above $1-4\times10^{-3}$ for $g_0\tau=0.5$. Additionally, the function is extremely flat in the wide range of values between $g_0\tau=0$ and $g_0\tau=0.5$. Therefore the scheme would be insensitive to variations in the interaction times within this range.

For the scheme to be useful, the probability of a single run leading to the successful generation of an entangled state, $P_{\text{success}}$, needs to be high enough that prohibitively many repetitions are not required.
\begin{equation}
P_{\text{success}}=1-\cos^4(g_0\tau)=2(g_0\tau)^2+O\bigl((g_0\tau)^4\bigr)
\end{equation}
The success probability is plotted in figure \ref{fig:idealprob}. For $g_0\tau=0.5$, this probability is approximately 0.4, so a successful run would probably be achieved in 2 or 3 repetitions.
The optimal parameter range for the scheme depends upon the fidelity of the state required. The higher the value of $g_0\tau$ chosen the higher the success probability will be but the lower the fidelity of the state generated. If, for example, fidelities of $0.95$ were acceptable, and a minimum success probability of 0.5 were desired, the scheme could operate between approximately $g_0\tau=0.6$ and $g_0\tau=0.9$. The exact value of $g_0\tau$, however, can lie anywhere in this range, so no fine tuning of the interaction time is required.
Note that even if extremely high fidelities such as 0.999 are required, the necessary parameters (up to $g_0\tau=0.35$), still allow a success probability of up to 0.22,  meaning that a successful run would probably be reached after 4 or 5 repetitions.

\section{Practical Considerations}\label{practical}

\subsection{Implementation}\label{implementation}

The above description is quite general and  no
particular type of atom or cavity has been specified .
A number of aspects must be considered in the choice of a physical
system to implement the scheme.
Firstly, it would be desirable that the entangled states, once
generated, would be as long-lived as possible. This favors
microwave cavities over optical cavities, since the lifetime of a
photon in an optical cavity is currently at the very most one
microsecond \cite{WolfgangLange}, whereas microwave cavities with
a photon lifetime of a millisecond have been made
\cite{harochermp}.
It takes an atom traveling at 500ms$^{-1}$ around 20$\mu$s to
traverse a typical microwave cavity, which is much smaller than
this lifetime, so there would be sufficient time for further atoms
to probe and interact with the cavity mode before the entangled
state has dissipated.
If a microwave cavity is used, an atom with a microwave transition
is then needed. Microwave transitions occur in atomic fine and
hyperfine structure, however dipole transitions between these
states are forbidden, and their interaction with the cavity mode
would be much too weak to implement this scheme -- the interaction
times that would be required would be much greater than the photon
lifetime of the cavity.
Rydberg atoms, on the other hand, although more difficult to
prepare, have large dipole moments and thus would interact
strongly with cavity modes.

Experiments have been carried out, for example by the Paris group
of Haroche and coworkers \cite{harochermp}, which have parameters
close to that required in this scheme. Recall that  our scheme
requires that the product of parameters $g_0$ and $\tau$ is at the
minimum 0.2 and maximally 0.5 to 0.8, depending on the fidelity of
Bell state one wants to generate.
In the Paris experiments, a Rydberg atom interacts with a
microwave cavity Rydberg atoms interact resonantly with microwave
cavity modes with a Rabi frequency of approximately 47kHz. This
means that  $g_0=47000\pi s^{-1}=1.48\times10^{5}
s^{-1}$.
Atoms in an atomic beam from an oven source travel at speeds of
the order of hundreds of meters per second. In the Paris
experiment atoms with a speed of 500ms$^{-1}$ are selected. This
means that the effective interaction time is such that a single
Rabi oscillation is performed, i.e. $g_0\tau=\pi$.
This is a factor of 4 to 8 lower than the parameter range for our
scheme. It would be difficult to lower $\tau$ by using faster
atoms, since the velocity of the atoms scales with the square root
of the atom oven temperature, so a lower $g_0$ would be required.
This could be obtained be using a larger cavity, and since $g_0$
scales with $1/\sqrt{V}=1/L^{\frac{3}{2}}$, so a cavity of mirror
separation 3 or 4 times as great as in the Paris experiments would
lead to a effective interaction times in the required range. Thus
Rydberg atoms and microwave cavities could be employed to
implement the scheme.

The disadvantage in using Rydberg atoms is that, at present, the efficiency of state detection schemes is low. In  \cite{harochermp}, for example, they report a detection efficiency of 40\%. We will discuss the implications this has for the scheme in subsection \ref{detection}.
First, however, we consider the effect on the fidelity of the
entangled states produced in the scheme if the path of the atom
through the cavities is not well controlled and deviates from the
line through the center of the cavities.

\subsection{The Atomic Path}\label{collim}

If the effective interaction times with both cavities are not exactly the same, this can reduce the fidelity of the entangled state produced by a successful run of the scheme.
 In our discussion above, we assumed that the two interaction times were exactly equal. In practice, however, it could be difficult to control the path of the atom so precisely.
Let us consider first the effect that differing interaction times
would have in the fidelity of the Bell state generated by a
successful run of the scheme. Let the effective interaction
between the atom and cavity A and the atom and cavity B be $\tau$
and $\tau(1-\epsilon)$. We can rewrite equation
\eqref{eq:idealfid}, to take the differing interaction times into
account, and find the following expression for the fidelity of the
entangled state generated by a successful run of the scheme.
\begin{equation}
F=\frac{1}{2}+\frac{\cos(g_0\tau)\sin(g_0\tau)\sin\bigl(g_0\tau(1-\epsilon)\bigr)}
{\cos^2(g_0\tau)\sin^2\bigl(g_0\tau(1-\epsilon)\bigr)+\sin^2(g_0\tau)}
\end{equation}

\begin{figure}
\includegraphics{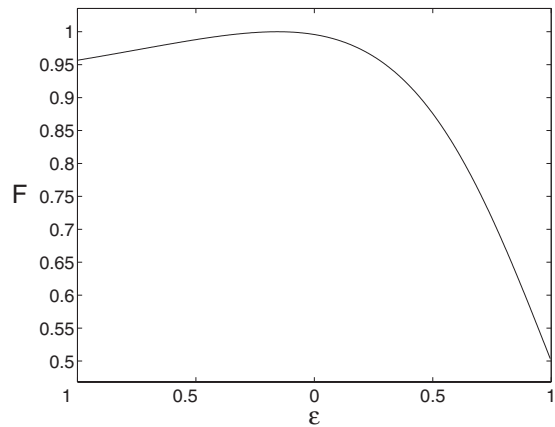}
\caption{\label{fig:epsilon} The fidelity of the Bell state generated in the cavity, if the effective interaction times are $\tau$ and $\tau(1-\epsilon)$ for the interactions with cavities A and B respectively, is plotted here as a function of $\epsilon$ for $g_0\tau=0.5$. }
\end{figure}

\begin{figure}
\includegraphics{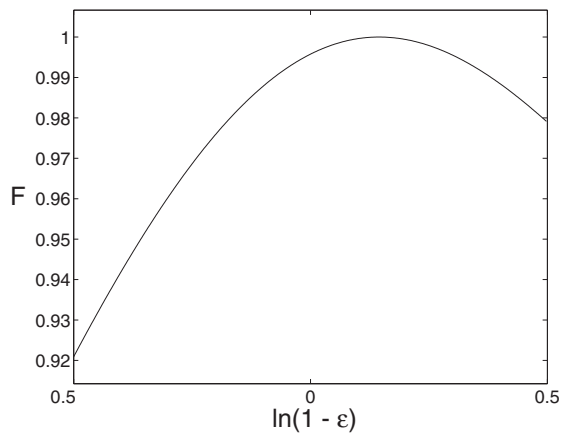}
\caption{\label{fig:logepsilon} The fidelity of the Bell state generated in the cavity plotted again as a function of $\epsilon$ for $g_0\tau=0.5$, this time plotted against $\ln(1-\epsilon)$. }
\end{figure}

The fidelity is plotted for $g_0\tau=0.5$ as a function of $\epsilon$ in figure \ref{fig:epsilon}.
The asymmetry of the plot is partly an artifact of the choice of parameterization, but, if we take this into account, by plotting $F$ again $\ln(1-\epsilon)$, as in figure \ref{fig:logepsilon}, we see that the asymmetry remains. This is due to the asymmetry in the scheme itself regarding the interactions with the two cavities. When the atom enters the first cavity it is always in the product state $\ket{e}$, whereas, when entering the second it is always entangled with the first cavity. This makes the interaction with cavity $B$ in some sense, slightly weaker, and is the reason that the maximal value of $F$ occurs when $\epsilon$ has a small negative value. The slightly longer interaction time compensates for the interaction being slightly weaker.

Since it is reasonably easy to position the cavities in the
desired place to very high precision, let us assume that they are
perfectly aligned in the layout illustrated in figure
\ref{fig:layout}.
\begin{figure}
\includegraphics[scale=0.35]{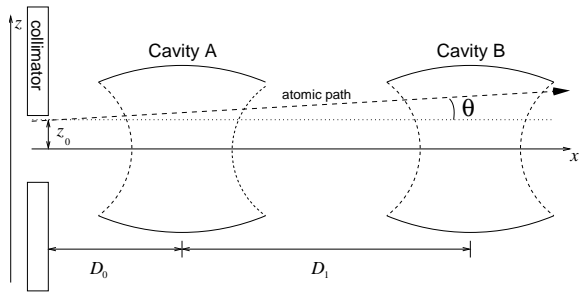}
\caption{\label{fig:zdiag} A general straight line path through the cavities can be defined in terms of the parameters $y_0$, $z_0$, $\phi$ and $\theta$, where  distances are measured in meters and  angles in radians. This figure illustrates $z_0$ and $\theta$, the other two parameters, $y_0$ and $\phi$, are equivalently defined in the perpendicular $x-y$-plane.
The distance between the exit of the collimator and the center of cavity A is $D_0$. The distance between the centers of the cavities is $D_1$.
}
\end{figure}
The factor which would be harder to control would be the atom's path through the apparatus, since it would be traveling on a ballistic path after being ejected from a heat oven, and although such atomic beams can be highly collimated, there will generally be a small amount of spread in the transverse direction, meaning that the atom's path will diverge slightly from the central path.

To calculate how much this would affect the interaction times we consider the cavity geometry.
The spherical mirrors commonly used in cavity QED experiments support Gaussian cavity modes, which have the following mode function
\begin{equation}
u(x,y,z)=e^{-\frac{x^2+y^2}{w_0^2}}\cos{\frac{2\pi z}{\lambda}},
\end{equation}
where $w_0$ is the mode waist. The z-axis lies along the line connecting the centers of the two mirrors and the origin is in the center of the cavity. The waist of a Gaussian mode, $w_0$, is a function of the cavity geometry and the field wavelength,
\begin{equation}
w_0=\left[\frac{\lambda\sqrt{L}}{2\pi}\sqrt{R-L}\right]^{\frac{1}{2}},
\end{equation}
where $L$ is the separation between and $R$ the radius of curvature of the mirrors.
If the atomic travels along the central axis of the cavity with constant speed $v$, the effective interaction time is $\sqrt{\pi}w_0/v$.
Let us consider a general straight path through the system, which can be defined in terms of four parameters, $y_0$ and $z_0$, the initial displacements from the central line in the $y$ and $z$ directions, and $\phi$ and $\theta$ the angles between the atomic path and the central line in the  $y$ and $z$ directions, illustrated for $z_0$ and $\theta$ in figure \ref{fig:zdiag}.
We can calculate the effective interaction times between the atom and cavity as it travels along this path with speed $v$ and find, that the interaction time with each cavity, in terms of displacements $\delta_y$ and $\delta_z$ which differ for each cavity, is

\begin{equation}\label{effectint}
\begin{split}
\tau_{{\rm
eff}}=&\frac{\sqrt{\pi}\omega_0}{v\cos{\theta}}\exp\left[-\frac{\delta_y^2}{\omega_0^2}\left(1-\frac{\sin^2\theta}{\cos^2\phi}\right)\right]\exp\left[-\frac{k^2\omega_0^2\tan^2\theta}{4}\right]\\
&\cos\left[k\delta_z-k\delta_y\left(\frac{\sin\theta\sin\phi}{\cos^2\theta}\right)\right]
.
\end{split}
\end{equation}
For the interaction with cavity $A$, $\delta_y=y_0+\cos(\phi)D_0$
and $\delta_z=z_0+\cos(\theta)D_0$, and for the interaction with
cavity $B$ $\delta_y=y_0+\cos(\phi)(D_0+D_1)$ and
$\delta_z=z_0+\cos(\theta)(D_0+D_1)$.
We can use these expressions to calculate $\epsilon$ in terms of
these parameters, written here to the  second order in $y_0$,
$z_0$, $\phi$ and $\theta$ and their products,

\begin{equation}\label{eq:epsilon}
\begin{split}
\epsilon\approx&\frac{1}{w_0^2}\left[(D_1\phi)^2+(D_1\phi)(2D_0\phi)+(2y_0)(D_1\phi)\right]\\&\mbox{}+\frac{2\pi^2}{\lambda^2}\left[(D_1\theta)^2+(D_1\theta)(2 D_0\theta)+(2z_0)(D_1\theta)\right].
\end{split}
\end{equation}
Let us discuss the constraints this would have on the
collimation of a typical experiment. In the cavity QED experiments
of Haroche and coworkers \cite{harocheprl761800,harochermp},
Rydberg atoms interact resonantly with microwave cavities. In a
typical experiment, a cavity mode with waist $w_0=5.97$mm and
wavelength $\lambda=5.87$mm is employed. For $\epsilon$ to be
small, the quantities in parentheses in equation
\eqref{eq:epsilon} must be much smaller than $w_0$ and
$\lambda/\sqrt{2}\pi=1.32$mm. This means that the atomic beam must
be collimated so that the effective beam radius is much smaller
than this distance. In \cite{harocheprl761800} an effective  beam
radius of $0.25$mm is reported. If we assume from this that, in
the worst case, this would mean that $y_0,z_0\approx$0.25mm and
$D_1\phi,D_1\theta\approx$0.25mm, we can estimate that $\epsilon$
would be less than $0.2$. If $g_0\tau=0.8$ and $\epsilon=0.2$,
this would corresponds to a reduction of the fidelity of the
entangled state produced by a successful run from 0.96 if both
interaction times are exactly equal to 0.93.
Thus, with the beam collimation currently available in the
laboratory, Bell states with a fidelity high enough, for example,
to exhibit significant violations of the Bell inequalities
\cite{someone}, could be generated.

Equation \eqref{eq:epsilon} also provides another reason why
optical cavities would be unsuitable for the scheme. The typical
waist of an optical cavity  tends to be much smaller
than that of a microwave cavity , so the demands on
the atomic beam collimation required if optical cavities were used
would be extremely high.

\subsection{Detector Efficiency}\label{detection}
Our analysis in the previous section assumes that the atomic state
detector has perfect efficiency. In practice, the detection
efficiency will be less than unity. Indeed, as mentioned above,
current state detection methods for Rydberg atoms have an
efficiency of just 40\% \cite{harochermp}. The detection process
ionizes the atom, destroying the state, so increased efficiency
could not be obtained by placing detectors in series.
A single detection failure will disrupt the scheme, since it will
cause a mixed state to be created in the cavities. The scheme must
then be halted, and one would then have to wait until this state
would have dissipated from the cavities, and the cavities have
returned to the vacuum state. Otherwise, if further atoms are sent
through the cavities immediately, they will interact with the
mixed state, and any ``successful'' run, will have generated a
mixed state with much reduced fidelity. Rather than halting the
flow of atoms through the cavities, their interaction with the
cavities could be prevented for the cavity dissipation time by the
application of an electric field to the system, to create a Stark
shift in the atoms such that they are no longer resonant with the
cavities.
Therefore, one would like a detection efficiency  high enough that the probability of a detection failure, during the typical number of runs needed before a successful ground state measurement is made, is low.
The mean number of runs to generate the entangled state in the cavity, if the detector were ideal, is simply the inverse of the success probability. The probability that the detector works every time during the process is therefore simply $D^{1/P_{\text{success}}}$, which for all values of $P_{\text{success}}$ is less than or equal to $D$. Therefore, for a reliable scheme, a high detection efficiency would be desirable.

For example, let us consider an implementation of this scheme with Rydberg atoms traveling through the cavities such that $g_0\tau=0.5$. The success probability  $P_{\text{success}}$ is 40.7\% and with current detectors with efficiency 40\%, $P_{\text{det}}$ would be around 10\%. This would  mean, typically one would have to repeat the whole process, including preparation of the cavities, 10 times before it could reach its successful conclusion. However, in light of the fact that the cavity dissipation time is of the order of milliseconds, and the time taken for each run much less than this, even in this case, the time needed to repeat the scheme enough times to generate the Bell state would be a fraction of a second.

\section{Conclusion}

We have proposed a scheme for the generation of high-fidelity Bell
states between two spatially separated cavity modes. The scheme is
non-deterministic, but we have shown that within the range of
parameters $g_0\tau$ between approximately 0.3 and 0.9, fidelities
higher than 0.95 are obtained, with success probabilities for a
single run greater than 1/5 for this entire range.

The most appropriate physical system to implement this scheme
would be a combination of microwave cavities and Rydberg atoms.
The low detection efficiency for Rydberg states would increase the
number of times which the scheme would need to be repeated before
a successful run, and require extra time after each detection
failure to allow the mixed state produced in the cavities to
dissipate. Nevertheless, the scheme would still be successful
within a reasonable number of repetitions. The scheme requires
that the atomic beam used is highly collimated otherwise the
fidelity of the states produced may b degraded, but the
collimation which has already been achieved in current experiments
is high enough that this effect would be small.

The principle of using brief interactions and measurement to
generate entanglement, could be adapted to many other physical
systems, and may be especially useful in systems where interaction
times are hard to control.

\begin{acknowledgments}
This work was supported by EPSRC, the EQUIP project of the
European Union, European Science Foundation program on ``Quantum
Information theory and Quantum Computing'', Hewlett-Packard Ltd.
and US-Army grant DAAD19-02-1-0161.

\end{acknowledgments}

{\em Note added in proof:} After this work was completed, we became aware of \cite{messina}, in which the author presents a different scheme for the generation of Bell states in two cavities via the passage of a single atom through the cavities.

\end{document}